\documentclass[a4paper,aps,prd,10pt,preprintnumbers,showpacs,twocolumn,superscriptaddress,nofootinbib,nobibnotes,floatfix,amsmath,amssymb]{revtex4-1}
\usepackage{graphicx}
\usepackage{cmap}
\usepackage[utf8]{inputenc}
\usepackage[T1]{fontenc}

\newcommand{\re}[1]{\operatorname{Re}\!\left(#1\right)}
\newcommand{\im}[1]{\operatorname{Im}\!\left(#1\right)}
\newcommand{\calK}{\mathcal{K}}
\newcommand{\Order}[1]{\mathcal{O}\!\left(#1\right)}

\begin{document}

\title{Grey-Body Factors and Thermodynamics of Asymptotically de Sitter Black Holes in Generalized Proca Theory}
\author{Bekir Can L{\"u}tf{\"u}o{\u{g}}lu}
\email{bekir.lutfuoglu@uhk.cz}
\affiliation{Department of Physics, Faculty of Science, University of Hradec Kr{\'a}lov{\'e}, Rokitansk{\'e}ho 62/26, 500 03 Hradec Kr{\'a}lov{\'e}, Czech Republic}


\author{Javlon~Rayimbaev}
\email{javlon@astrin.uz}
\affiliation{Institute of Theoretical Physics, National University of Uzbekistan, Tashkent 100174, Uzbekistan}
\affiliation{Kimyo International University in Tashkent, Shota Rustaveli Street 156, Tashkent 100121, Uzbekistan}

\author{Nuriddin Kurbonov}
\email{n.kurbonov@newuu.uz}
\affiliation{Ulugh Beg Astronomical Institute, Astronomy St. 33, Tashkent 100052, Uzbekistan}
\affiliation{University of Tashkent for Applied Sciences, Gavhar Str. 1, Tashkent 700127, Uzbekistan}
\affiliation{Tashkent State Technical University, Tashkent 100095, Uzbekistan}

\author{Sardor~Murodov}
\email{mursardor@ifar.uz}
\affiliation{New Uzbekistan University, Movarounnahr Str. 1, Tashkent 100000, Uzbekistan}
\affiliation{Tashkent International University of Education, Imom Bukhoriy 6, Tashkent 100207, Uzbekistan}

\author{Faisal~Javed}
\email{faisaljaved.math@gmail.com }\affiliation{College of Transportation, Tongji University, Shanghai 201804, People's Republic of China}
\affiliation{Research Center of Astrophysics and Cosmology, Khazar University, Baku, AZ1096, 41 Mehseti Street, Azerbaijan}

\begin{abstract}
Generalized Proca theory supplements gravity with a massive vector field whose derivative self-interactions can support black holes carrying primary vector hair. In the asymptotically de Sitter branch considered here, the de Sitter scale is effective: it is generated by the vector sector rather than imposed through a bare cosmological-constant term.  We compute grey-body factors and effective absorption cross-sections for massive scalar, electromagnetic, and massless Dirac test fields on this background. The transmission curves are obtained from a sixth-order WKB barrier calculation and are compared with the quasinormal-mode reconstruction; the two descriptions mostly agree, with small visible differences at lower mutlipoles. Increasing the scalar mass raises and broadens the scalar barrier, suppresses transmission at fixed frequency, and shifts efficient transmission to higher frequencies. The couplings considerably affect the grey-body factors and black hole thermodynamics. 
\end{abstract}

\maketitle

\section{Introduction}

Grey-body factors complement quasinormal modes as probes of black-hole geometry. While quasinormal frequencies characterize the poles of the response function, the grey-body factors measure barrier penetration and therefore quantify how efficiently waves are transmitted across the effective potential separating the event and cosmological horizons. Their role in black-hole emission calculations is well established in the literature~\cite{Page:1976df,Page:1976ki,Kanti:2004nr}. In Hawking-radiation spectra~\cite{Hawking:1975vcx}, this filtering can be as important as the temperature itself, because the observable flux is controlled by the combined effect of thermal occupation and transmission through the curvature barrier~\cite{Konoplya:2019ppy}. In this sense grey-body factors provide information that is directly relevant both for classical scattering and for quantum emission.

The dynamics of test-field perturbations in asymptotically de Sitter spacetimes is also interesting from a broader perspective. The quasinormal spectrum fixes the characteristic relaxation time of the static patch, controls the late-time approach to equilibrium, and, in charged de Sitter geometries with an inner horizon, enters the Strong Cosmic Censorship bound through the comparison between the spectral gap and the Cauchy-horizon surface gravity. Perturbations and spectra of de Sitter black holes have therefore been studied extensively for different spins, dimensions, and higher-curvature extensions~\cite{Skvortsova:2023zmj,Konoplya:2017lhs,Mo:2018nnu,Skvortsova:2023zca,Konoplya:2004uk,Dyatlov:2010hq,Jansen:2017oag,Konoplya:2014lha,Konoplya:2007jv,Molina:2003ff,Zhidenko:2003wq,Aragon:2020qdc,Konoplya:2007zx,Jing:2003wq,Kanti:2005xa,Konoplya:2017ymp,Konoplya:2013sba,Lutfuoglu:2025hjy}.

For asymptotically de Sitter black holes, the scattering problem is naturally formulated in the static patch, where the event horizon and the cosmological horizon play symmetric roles as the two boundaries of the radial domain. This makes the generalized Proca black-hole family especially attractive, because the same Proca sector that supports primary hair also generates the effective cosmological scale of the asymptotics~\cite{Heisenberg:2014rta,Charmousis:2025jpx,RefProcaGB2026}. 

For generalized Proca theory itself, the interest goes beyond a particular black-hole solution: the derivative self-interactions of a massive vector can be organized so that only the three physical vector polarizations propagate, while beyond-generalized extensions preserve this property through degeneracy conditions~\cite{Heisenberg:2014rta,Heisenberg:2016BeyondGP}. Cosmological and screening analyses also provide a useful stability language, because viable branches require positive kinetic coefficients and positive squared sound speeds in the tensor, vector, and scalar sectors~\cite{DeFelice:2016CosmoGP,DeFelice:2016ScreenGP}.

The scattering problem is linked to the problem of finding proper oscillation frequencies of black holes, called quasinormal modes \cite{Kokkotas:1999bd,Berti:2009kk,Konoplya:2011qq,Bolokhov:2025rng}. Recent studies of the same geometry have already shown that the quasinormal spectra of massless and massive test fields are highly sensitive to the Proca-hair parameter and to the couplings controlling the effective de Sitter radius \cite{Bolokhov:2026dfg,Skvortsova:2026idf}. This same setting also motivates a dedicated grey-body-factor analysis.

Perturbations, optical phenomena and quasinormal modes of asymptotically flat black holes in the presence of Proca hairs have been studied in \cite{Lutfuoglu:2025qkt,Konoplya:2025bte,Konoplya:2025uiq}.
A quasinormal-mode analysis of massless scalar, electromagnetic, and Dirac perturbations on the same effective-de Sitter generalized Proca geometry has also shown that the resulting spectra provide useful input for grey-body factors~\cite{Skvortsova:2026gptqnm}.

We compare three test-field sectors on the generalized Proca background: with an effective de Sitter asymptotic \cite{Charmousis:2025jpx,RefProcaGB2026} a massive scalar field, an electromagnetic field, and a massless Dirac field. Their effective potentials differ in a specific way: the massive scalar sector contains both the curvature contribution $f'(r)/r$ and the extra scale $\mu$, the electromagnetic barrier is purely centrifugal, and the Dirac problem is governed by supersymmetric partner potentials. Comparing the corresponding transmission probabilities then isolates which features come from the background geometry and which arise from the field spin or mass.

The paper is organized as follows. In Sec.~II we summarize the generalized Proca black-hole metric and its effective de Sitter asymptotics. Section~III collects the master equations for the massive scalar, electromagnetic, and Dirac fields relevant for the grey-body problem. In Sec.~IV we specify the scattering boundary conditions and define the transmission probabilities. Section~V introduces the corresponding effective absorption cross-sections in the static patch. Section~VI discusses the temperature choices, and Sec.~VII summarizes the main conclusions.

\section{Generalized Proca Black-Hole Geometry}

We consider the static and spherically symmetric generalized Proca black holes discussed in Refs.~\cite{Heisenberg:2014rta,Charmousis:2025jpx,RefProcaGB2026}. The line element is written as
\begin{equation}
\mathrm{d}s^2=-f(r)\,\mathrm{d}t^2+\frac{\mathrm{d}r^2}{f(r)}+r^2\mathrm{d}\Omega_2^2,
\end{equation}
with metric function
\begin{equation}
\begin{split}
 f(r)=&\,1-\frac{2(M-Q)}{r}
 +\frac{r^2}{2\alpha}\Bigg(A \\
 &-\sqrt{B+\frac{8\alpha}{r^3}\left[Q+\frac{\lambda(M-Q)}{2\beta}\right]}\Bigg),
\end{split}
\label{eq:metric_function_gbf}
\end{equation}
where
\begin{equation}
A\equiv 1-\frac{\beta\lambda}{2},
\qquad
B\equiv 1-\beta\lambda\left(1-\frac{c_1\lambda}{4}\right).
\end{equation}
Here, $M$ is the mass-type integration constant and $Q$ is the primary-hair parameter associated with the Proca sector.

For the asymptotically de Sitter branch, the static patch lies between the event horizon $r_h$ and the cosmological horizon $r_c$, namely between the two positive outer roots of $f(r)=0$. It is convenient to introduce
\begin{equation}
\Lambda_{\rm eff}\equiv \frac{3}{2\alpha}\left(\sqrt{B}-A\right),
\qquad
H\equiv \sqrt{\frac{\Lambda_{\rm eff}}{3}},
\end{equation}
so that $H$ serves as the effective Hubble parameter, dynamically set by the Proca couplings. The tortoise coordinate is defined in the usual way,
\begin{equation}
\frac{\mathrm{d}r_*}{\mathrm{d}r}=\frac{1}{f(r)},
\label{eq:tortoise_gbf}
\end{equation}
with $r_*\to-\infty$ at the event horizon and $r_*\to+\infty$ at the cosmological horizon.

\section{Massive Scalar, Electromagnetic, and Dirac Test Fields}

For all sectors considered below, the radial problem can be reduced to a Schr\"odinger-type equation,
\begin{equation}
\frac{\mathrm{d}^2\Psi}{\mathrm{d}r_*^2}+\left[\omega^2-V(r)\right]\Psi=0,
\label{eq:master_generic_gbf}
\end{equation}
where the spin dependence is encoded entirely in the effective potential.

\subsection{Massive scalar field}

The minimally coupled scalar of mass $\mu$ satisfies
\begin{equation}
\left(\Box-\mu^2\right)\Phi=0.
\end{equation}
Using the standard decomposition~\cite{Carter:1968ks,Konoplya:2018arm}
\begin{equation}
\Phi=e^{-i\omega t}Y_{\ell m}(\theta,\varphi)\frac{U(r)}{r},
\end{equation}
one finds the radial equation
\begin{equation}
\frac{\mathrm{d}^2U}{\mathrm{d}r_*^2}+\left[\omega^2-V_{\mu}(r)\right]U=0,
\end{equation}
with effective potential~\cite{Konoplya:2018arm}
\begin{equation}
V_{\mu}(r)=f(r)\left[\frac{\ell(\ell+1)}{r^2}+\frac{f'(r)}{r}+\mu^2\right].
\label{eq:massive_scalar_potential_gbf}
\end{equation}
Compared with the two massless sectors, the scalar potential contains the curvature term $f'(r)/r$, while the mass term further raises the barrier and introduces an extra scale that can qualitatively alter the transmission probability.

\subsection{Electromagnetic field}

For the Maxwell field,
\begin{equation}
\nabla_\mu F^{\mu\nu}=0,
\qquad
F_{\mu\nu}=\partial_\mu A_\nu-\partial_\nu A_\mu,
\end{equation}
the master variable obeys
\begin{equation}
\left(-\frac{\partial^2}{\partial t^2}+\frac{\partial^2}{\partial r_*^2}-V_1(r)\right)\Psi^{(1)}_{\ell m}(t,r)=0,
\end{equation}
where
\begin{equation}
V_1(r)=f(r)\frac{\ell(\ell+1)}{r^2},
\qquad \ell\geq 1.
\label{eq:em_potential_gbf}
\end{equation}

\subsection{Massless Dirac field}

For the massless Dirac field the separated radial combinations satisfy~\cite{Jing:2003wq,LopezOrtega:2007dirac}
\begin{equation}
\left(-\frac{\partial^2}{\partial t^2}+\frac{\partial^2}{\partial r_*^2}-V_{\pm}(r)\right)Z_{\pm}(t,r)=0,
\end{equation}
with supersymmetric partner potentials
\begin{equation}
V_{\pm}(r)=W^2(r)\pm \frac{\mathrm{d}W}{\mathrm{d}r_*},
\qquad
W(r)=\kappa\frac{\sqrt{f(r)}}{r},
\label{eq:dirac_potential_gbf}
\end{equation}
where $\kappa=1,2,\ldots$. Under the standard black-hole boundary conditions, the two partner potentials are isospectral, so either one may be used in the numerical scattering problem.

\begin{figure*}[t]
\centering
\includegraphics[width=0.98\textwidth]{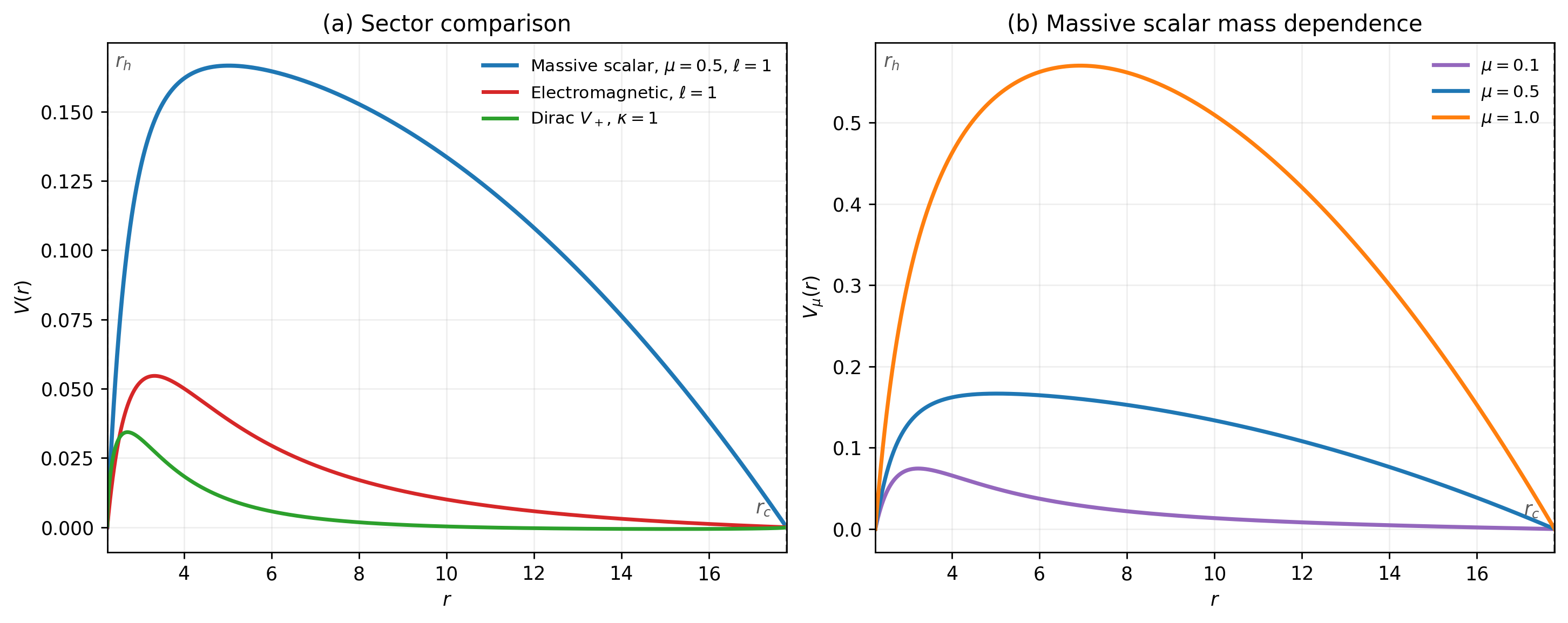}
\caption{Representative effective potentials for the de Sitter configuration $(\alpha,\beta,\lambda,c_1,Q,M)=(1,1,0.2,2,0,1)$. Left: comparison of the massive scalar potential for $(\mu,\ell)=(0.5,1)$, the electromagnetic potential for $\ell=1$, and the Dirac partner potential $V_+$ for $\kappa=1$. Right: massive scalar potential for $\ell=1$ and $\mu=0.1,0.5,1.0$. In all cases, the potential vanishes at the event horizon $r_h\simeq 2.248$ and the cosmological horizon $r_c\simeq 17.776$.}%
\label{fig:gbf-potentials}
\end{figure*}

To make the qualitative differences between the three sectors explicit, it is useful to inspect a representative asymptotically de Sitter configuration. In Fig.~\ref{fig:gbf-potentials} we choose
\begin{equation}
(\alpha,\beta,\lambda,c_1,Q,M)=(1,1,0.2,2,0,1),
\end{equation}
for which the event and cosmological horizons are located at
\begin{equation}
r_h\simeq 2.248,
\qquad
r_c\simeq 17.776.
\end{equation}
The left panel compares the massive scalar potential for $(\mu,\ell)=(0.5,1)$ with the electromagnetic barrier for $\ell=1$ and the Dirac partner potential $V_+$ for $\kappa=1$. Because $f(r)$ vanishes at both horizons, all three potentials go to zero at the ends of the static patch, so the grey-body problem is governed by transmission through a single barrier in the interval $(r_h,r_c)$. For these representative choices, the massive scalar barrier is the highest and broadest one, with a maximum $V_{\mu}^{\max}\simeq 0.167$ near $r\simeq 5.03$. By comparison, the electromagnetic peak is $V_1^{\max}\simeq 5.46\times 10^{-2}$ near $r\simeq 3.33$, while the plotted Dirac partner reaches only $V_+^{\max}\simeq 3.43\times 10^{-2}$ near $r\simeq 2.71$.

The right panel shows how the massive scalar potential varies as the field mass is increased, with $\ell=1$ fixed. As $\mu$ grows from $0.1$ to $1.0$, the barrier is lifted substantially and its maximum moves outward from roughly $r\simeq 3.23$ to $r\simeq 6.93$. This trend is directly relevant to the grey-body factors: at fixed frequency, a higher, wider barrier suppresses transmission and shifts the onset of efficient transmission to larger $\omega$. The electromagnetic and Dirac sectors, in contrast, depend only on the geometry and the angular structure of the perturbation equations, so they provide the cleanest reference curves for separating genuine mass effects from the background dependence common to all three sectors.

\section{Grey-Body Factors in the Static Patch}

For real frequencies $\omega>0$, the scattering problem is defined by choosing a wave incident from the cosmological-horizon side of the static patch. In terms of the tortoise coordinate, the asymptotic behavior is written as
\begin{equation}
\Psi\sim T_{\omega}\,e^{-i\omega r_*},
\qquad r\to r_h \quad (r_*\to-\infty),
\label{eq:bh_boundary_gbf}
\end{equation}
and
\begin{equation}
\Psi\sim e^{-i\omega r_*}+R_{\omega}\,e^{+i\omega r_*},
\qquad r\to r_c \quad (r_*\to+\infty).
\label{eq:cosmo_boundary_gbf}
\end{equation}
Here $T_{\omega}$ and $R_{\omega}$ are the transmission and reflection amplitudes, respectively. The grey-body factor is then defined as
\begin{equation}
\Gamma(\omega)=|T_{\omega}|^2,
\qquad
\mathcal{R}(\omega)=|R_{\omega}|^2.
\end{equation}
For real frequencies and properly normalized fluxes, one has the usual conservation law
\begin{equation}
\Gamma(\omega)+\mathcal{R}(\omega)=1.
\label{eq:flux_conservation_gbf}
\end{equation}

The transmission problem is evaluated with two complementary semiclassical constructions. The first is the sixth-order WKB treatment of the barrier between $r_h$ and $r_c$, applied directly to the potential appearing in Eq.~\eqref{eq:master_generic_gbf}.  The second one uses the recently developed relation between grey-body factors and quasinormal frequencies, including corrections beyond the leading eikonal approximation~\cite{Konoplya:2024lir,Konoplya:2024vuj}. This QNM--GBF correspondence has been tested and applied in numerous black-hole settings~\cite{Malik:2024cgb,Lutfuoglu:2025mqa,Skvortsova:2024msa,Konoplya:2010vz,Dubinsky:2024vbn,Lutfuoglu:2026uzy,Bolokhov:2026eqf,Malik:2025dxn,Bolokhov:2024otn,Lutfuoglu:2025ldc,Malik:2024wvs,Lutfuoglu:2025eik,Bolokhov:2025egl,Bolokhov:2026uol,Bolokhov:2026kqu}. The assumptions behind it are most secure when the effective potential has one smooth dominant maximum and no additional trapping region. If the potential changes topology, for instance by passing into a double-well or double-barrier configuration~\cite{Konoplya:2025hgp}, the correspondence no longer encodes the scattering problem with sufficient accuracy. A  similar warning applies in higher-curvature models, where strong deformations of the gravitational or centrifugal sector can produce eikonal and catastrophic instabilities~\cite{Konoplya:2017zwo,Takahashi:2010gz,Dotti:2004sh,Konoplya:2017lhs,Konoplya:2017ymp}. In such regimes, reconstructing grey-body factors from quasinormal frequencies is expected to become unreliable, ambiguous, or to fail altogether. In this way, the direct scattering calculation and the resonance-based estimate probe the same potential barrier from two different sides: the former follows wave transmission at real $\omega$, while the latter reconstructs the transmission profile from the complex frequencies associated with the barrier peak.

In the WKB description, the result is written in terms of the quantity $\calK_n(\omega)$ determined by the local shape of the effective potential at its maximum. For each spin sector and each angular label $n$, we use
\begin{equation}\label{eq:wkbGamma}
\Gamma_n^{(s)}(\omega)=|T_n^{(s)}(\omega)|^2
=\frac{1}{1+e^{2\pi i\calK_n^{(s)}(\omega)}}.
\end{equation}
Here $\calK_n^{(s)}$ contains the derivatives of $V(r)$ with respect to the tortoise coordinate, evaluated at the top of the barrier, with the higher-order WKB corrections included up to sixth order. This approximation is especially well adapted to the present static-patch problem because the representative scalar, electromagnetic, and Dirac potentials vanish at both horizons and form a  smooth barrier \cite{Lutfuoglu:2026rqe,Lutfuoglu:2025kqp,Lutfuoglu:2025blw,Lutfuoglu:2025ohb,Lutfuoglu:2025ljm,Bolokhov:2025aqy,Pathrikar:2025gzu,Konoplya:2023moy,Skvortsova:2024eqi,
Skvortsova:2024atk,Bolokhov:2023ruj,Lutfuoglu:2025pzi,
Skvortsova:2024wly,
Bolokhov:2023bwm,Bolokhov:2024bke}. As usual, the method is most reliable away from extremely low multipoles; nevertheless, it provides an efficient and uniform way of comparing large parameter scans. Similar WKB constructions have been widely used for black-hole scattering and quasinormal-mode calculations in a broad range of geometries~\cite{Konoplya:2001ji,Malik:2024iky,Guo:2020caw,
Konoplya:2021ube,Lutfuoglu:2026gis,
Eniceicu:2019npi,Konoplya:2009hv,
Fernando:2016ftj,
Karmakar:2023cwg,Abdalla:2005hu,
Malik:2025erb,Konoplya:2005sy,
Breton:2017hwe,Lutfuoglu:2026gey,
Konoplya:2006ar,Wongjun:2019ydo,
Momennia:2018hsm,Kokkotas:2010zd,
Stuchlik:2025mjj,
Malik:2024sxv,Ishihara:2008re,
Konoplya:2023ppx,Hamil:2024ppj,
Hamil:2024njs}.

The QNM-based estimate is used as an independent check on the same transmission curves. In the eikonal expansion, the WKB parameter entering Eq.~\eqref{eq:wkbGamma} can be expressed through the fundamental quasinormal frequency associated with the same angular sector. With our sign convention, this relation takes the schematic form
\begin{equation}\label{eq:gbf_qnm_eikonal}
i\calK_n^{(s)}(\omega)=\frac{\omega^2-\re{\omega_0}^2}
{4\,\re{\omega_0}\,\im{\omega_0}}+\Order{n^{-1}},
\end{equation}
where the subleading terms encode the beyond-eikonal corrections given explicitly in Ref.~\cite{Konoplya:2024vuj}.  Notice that the above relation implies that the modes of the Schwarzschild branch of modes must be taken, while the frequencies of the de Sitter branch \cite{Konoplya:2022xid,Konoplya:2022gjp} cannot be reproduced via the WKB method and cannot be used in the correspondence.

In practice, Eqs.~\eqref{eq:bh_boundary_gbf}--\eqref{eq:flux_conservation_gbf} are evaluated separately for the massive scalar, electromagnetic, and Dirac potentials in Eqs.~\eqref{eq:massive_scalar_potential_gbf},~\eqref{eq:em_potential_gbf}, and~\eqref{eq:dirac_potential_gbf}. The Figs. \ref{fig:gbf-em-alpha} and \ref{fig:gbf-em-beta} below show both the WKB transmission probability and the QNM-based reconstruction, while the difference panels identify the frequency intervals where the two approximations are most sensitive to spin, mass, and Proca-hair effects.

\begin{figure*}[t]
\centering
\includegraphics[width=0.49\textwidth]{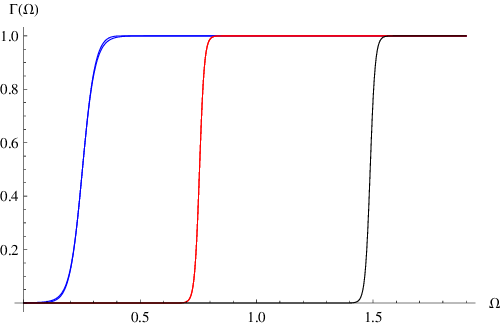}\hfill
\includegraphics[width=0.49\textwidth]{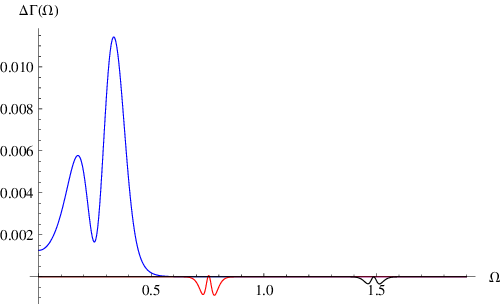}
\caption{Grey-body factors for the scalar $\ell=1$ mode in the neutral generalized Proca black hole background with $(M,Q,\alpha,\beta,\lambda,c_1)=(1,0,1,1,0.2,2)$. Left: transmission probability $\Gamma(\Omega)$ obtained by the sixth-order WKB method and by the correspondence with quasinormal modes. Right: difference $\Delta\Gamma(\Omega)$ between the two methods. The color coding is $\mu=0$ (blue), $\mu=1$ (red), and $\mu=2$ (black). For $\mu=1$ and $\mu=2$ the two determinations nearly overlap on the scale of the left panel.}%
\label{fig:gbf-wkb-qnm}
\end{figure*}

\begin{figure*}[t]
\centering
\includegraphics[width=0.49\textwidth]{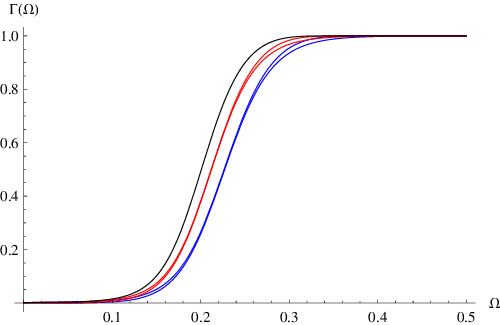}\hfill
\includegraphics[width=0.49\textwidth]{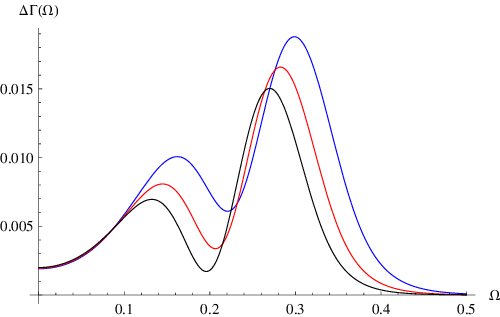}
\caption{Grey-body factors for the electromagnetic $\ell=1$ mode in the neutral generalized Proca black hole background with $(M,Q,\alpha,\beta,\lambda,c_1)=(1,0,1,1,0.2,c_1)$. Left: transmission probability $\Gamma(\Omega)$ obtained by the sixth-order WKB method and by the correspondence with quasinormal modes. Right: difference $\Delta\Gamma(\Omega)$ between the two methods. The color coding is $c_1=1.1$ (blue), $c_1=2.5$ (red), and $c_1=3.5$ (black).}%
\label{fig:gbf-em-c1}
\end{figure*}

\begin{figure*}[t]
\centering
\includegraphics[width=0.45\textwidth]{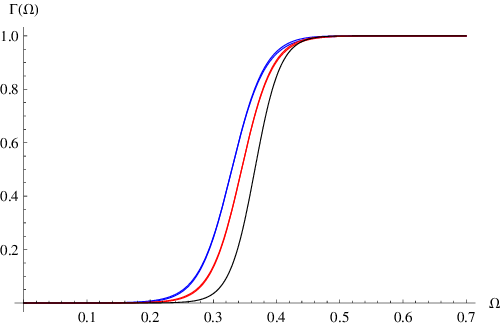}\hfill
\includegraphics[width=0.45\textwidth]{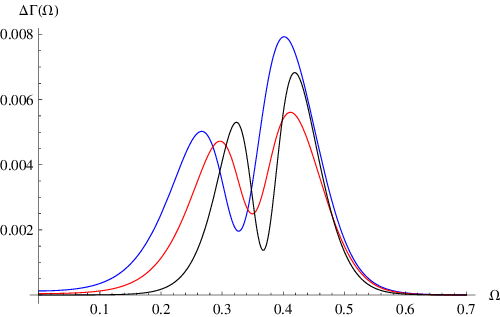}
\caption{Grey-body factors for the Dirac $\ell=3/2$ mode in the generalized Proca black hole background with $(M,Q,\alpha,\beta,\lambda,c_1)=(1,Q,1,1,0.2,2)$. Left: transmission probability $\Gamma(\Omega)$ obtained by the sixth-order WKB method and by the correspondence with quasinormal modes. Right: difference $\Delta\Gamma(\Omega)$ between the two methods. The color coding is $Q=0$ (blue), $Q=0.7$ (red), and $Q=1.03$ (black).}%
\label{fig:gbf-dirac-q}
\end{figure*}

The five comparisons in Figs.~\ref{fig:gbf-wkb-qnm}--\ref{fig:gbf-dirac-q} are best read together with the effective-potential picture of Fig.~\ref{fig:gbf-potentials}. In all cases, the qualitative rule is the standard one for barrier penetration: a higher and/or broader effective-potential peak lowers the grey-body factor at fixed frequency and shifts the onset of efficient transmission to larger $\omega$, whereas a lower barrier shifts the curve to the left. Figure~\ref{fig:gbf-wkb-qnm} shows this most directly for the scalar $\ell=1$ sector with $(M,Q,\alpha,\beta,\lambda,c_1)=(1,0,1,1,0.2,2)$. As already indicated by the representative potentials, increasing the field mass raises the scalar barrier, so the massless case $\mu=0$ reaches sizable transmission first, while the heavier cases $\mu=1$ and $\mu=2$ require larger frequencies to overcome the barrier. The right panel reflects the same tendency: the disagreement between the sixth-order WKB grey-body factor and the QNM-based estimate is most visible near the transmission threshold of the lightest field, whereas for $\mu=1$ and $\mu=2$ the two determinations are nearly indistinguishable on the plotting scale.

Within the electromagnetic sector, Fig.~\ref{fig:gbf-em-alpha} shows that increasing $\alpha$ hardens the barrier when $(M,Q,\beta,\lambda,c_1)=(1,0,1,0.2,2)$ is kept fixed. The $\alpha=0.1$ configuration is therefore the easiest to penetrate and reaches near-unit transmission already at rather low frequencies, whereas the $\alpha=1$ and $\alpha=1.9$ barriers are effectively higher and shift the rise of $\Gamma(\Omega)$ to larger $\Omega$. The right panel follows the same pattern: as the barrier becomes harder, the nonzero values of $\Delta\Gamma(\Omega)$ are shifted to a somewhat broader interval at higher frequencies.

\begin{figure*}[t]
\centering
\includegraphics[width=0.49\textwidth]{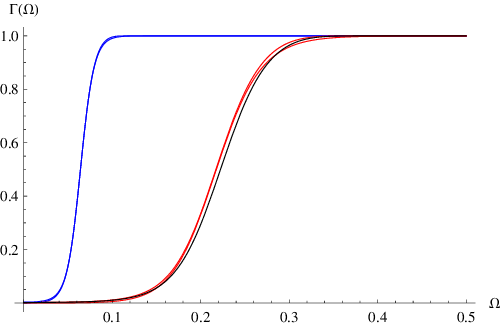}\hfill
\includegraphics[width=0.49\textwidth]{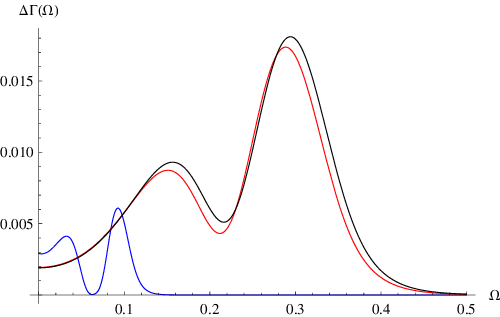}
\caption{Grey-body factors for the electromagnetic $\ell=1$ mode in the neutral generalized Proca black hole background with $(M,Q,\alpha,\beta,\lambda,c_1)=(1,0,\alpha,1,0.2,2)$. Left: transmission probability $\Gamma(\Omega)$ obtained by the sixth-order WKB method and by the correspondence with quasinormal modes. Right: difference $\Delta\Gamma(\Omega)$ between the two methods. The color coding is $\alpha=0.1$ (blue), $\alpha=1$ (red), and $\alpha=1.9$ (black).}%
\label{fig:gbf-em-alpha}
\end{figure*}

Figure~\ref{fig:gbf-em-beta} exhibits an even clearer monotonic trend with $\beta$. At fixed $\alpha=1$ and $(M,Q,\lambda,c_1)=(1,0,0.2,2)$, increasing $\beta$ makes the electromagnetic barrier taller and/or broader, so the transmission window opens later: the curve for $\beta=0.1$ reaches the plateau first, the case $\beta=1$ is intermediate, and $\beta=1.9$ is the most suppressed at low and intermediate frequencies. Consistently, the peaks of $\Delta\Gamma(\Omega)$ move to higher $\Omega$ and the disagreement is spread over a wider interval as the barrier hardens.

\begin{figure*}[t]
\centering
\includegraphics[width=0.49\textwidth]{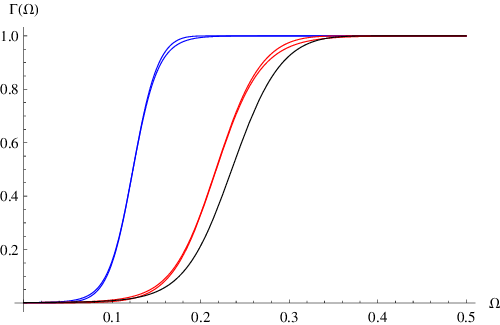}\hfill
\includegraphics[width=0.49\textwidth]{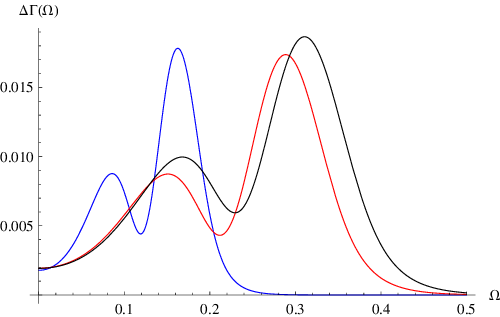}
\caption{Grey-body factors for the electromagnetic $\ell=1$ mode in the neutral generalized Proca black hole background with $(M,Q,\alpha,\beta,\lambda,c_1)=(1,0,1,\beta,0.2,2)$. Left: transmission probability $\Gamma(\Omega)$ obtained by the sixth-order WKB method and by the correspondence with quasinormal modes. Right: difference $\Delta\Gamma(\Omega)$ between the two methods. The color coding is $\beta=0.1$ (blue), $\beta=1$ (red), and $\beta=1.9$ (black).}%
\label{fig:gbf-em-beta}
\end{figure*}

Varying $c_1$ at fixed $(M,Q,\alpha,\beta,\lambda)=(1,0,1,1,0.2)$ reverses this ordering, because now the electromagnetic barrier is softened as $c_1$ increases. This is why Fig.~\ref{fig:gbf-em-c1} shows the largest transmission for $c_1=3.5$ and the smallest for $c_1=1.1$ at the same frequency: a lower barrier is penetrated more easily, so the onset of efficient transmission is shifted to smaller $\Omega$. The right panel mirrors the same softening, since the dominant maximum of $\Delta\Gamma(\Omega)$ moves to lower frequencies and decreases slightly in height as $c_1$ grows, while the dip between the two local maxima becomes more pronounced.

Finally, Fig.~\ref{fig:gbf-dirac-q} shows that increasing the primary-hair parameter $Q$ hardens the Dirac $\ell=3/2$ barrier for fixed $(M,\alpha,\beta,\lambda,c_1)=(1,1,1,0.2,2)$. The neutral case $Q=0$ is therefore the most transmissive, whereas $Q=0.7$ and especially $Q=1.03$ suppress the grey-body factor up to progressively higher frequencies. In the right panel, the discrepancy remains concentrated in the same barrier-top region, but its profile becomes more structured as $Q$ increases, culminating in the pronounced dip between two local maxima for $Q=1.03$.

\section{Absorption Cross-Sections}

In addition to the grey-body factor itself, it is often useful to repackage the same transmission information as an effective absorption cross-section. In asymptotically flat spacetimes, this quantity compares the absorbed flux with the incident plane-wave flux and may be interpreted as an effective capture area of the black hole. For asymptotically de Sitter black holes, there is no standard plane-wave region at spatial infinity, so in the present setting, the corresponding quantity should be understood as a generalized static-patch absorption cross-section: it measures how efficiently radiation sent inward from the cosmological horizon is transmitted through the potential barrier and captured by the event horizon.

For each sector, one may define a partial absorption cross-section for a given angular mode by factoring out the appropriate incident flux and angular multiplicity,
\begin{equation}
\sigma_n^{(s)}(\omega)=\frac{{\cal N}_n^{(s)}}{\omega^2}\,\Gamma_n^{(s)}(\omega),
\label{eq:partial_absorption_gbf}
\end{equation}
where $s$ labels the spin sector, $n$ denotes the angular quantum number ($n=\ell$ for the scalar and electromagnetic fields and $n=\kappa$ for the Dirac field), $\Gamma_n^{(s)}(\omega)$ is the corresponding grey-body factor, and ${\cal N}_n^{(s)}$ is the spin-dependent multiplicity factor. For the scalar field in four dimensions, one has ${\cal N}_\ell^{(0)}=\pi(2\ell+1)$, while the electromagnetic and Dirac sectors are treated analogously with the appropriate mode degeneracies. The total effective absorption cross-section is then obtained by summing over all propagating partial waves,
\begin{equation}
\sigma_{\rm abs}^{(s)}(\omega)=\sum_n \sigma_n^{(s)}(\omega).
\label{eq:total_absorption_gbf}
\end{equation}

Because Eqs.~\eqref{eq:partial_absorption_gbf} and~\eqref{eq:total_absorption_gbf} are built directly from the grey-body factors, they inherit the same barrier-penetration physics. Whenever the effective potential becomes higher or broader, the corresponding partial transmission probability decreases at fixed frequency, and the partial absorption cross-section is suppressed and shifted to larger $\omega$. Conversely, when the barrier is softened, the absorption sets in already at lower frequencies. Thus, the trends discussed above for the scalar, electromagnetic, and Dirac grey-body factors translate directly into the corresponding effective capture areas: harder potentials reduce the absorption efficiency, whereas softer potentials enhance it.

From a physical point of view, the absorption cross-section provides a complementary description of the same scattering process. At low frequencies, it quantifies how strongly long-wavelength radiation couples to the black hole despite the de Sitter barrier, while at higher frequencies, it approaches a more geometric measure of capture determined mainly by the background geometry. Since the effective absorption cross-sections are constructed directly from the grey-body factors, no additional dynamical input is required beyond the transmission probabilities themselves, and the same framework applies to the scalar, electromagnetic, and Dirac sectors.

Figure~\ref{fig:acs-em-charge-comparison} compares representative electromagnetic absorption cross-sections for the neutral configuration $(M,Q,\alpha,\beta,\lambda,c_1)=(1,0,1,1,0.2,2)$ and for the corresponding near-extreme charged configuration with $Q=1.02$. In both cases, the colored curves denote the partial absorption cross-sections for successive multipoles, while the black curve is the corresponding total effective absorption cross-section. The same multipole-by-multipole buildup is present in the two backgrounds, but the larger value of $Q$ visibly reduces both the dominant low-$\ell$ peaks and the high-frequency level around which the total cross-section oscillates.

\begin{widetext}
\refstepcounter{figure}\label{fig:acs-em-charge-comparison}
\begin{center}
\includegraphics[width=0.415\textwidth]{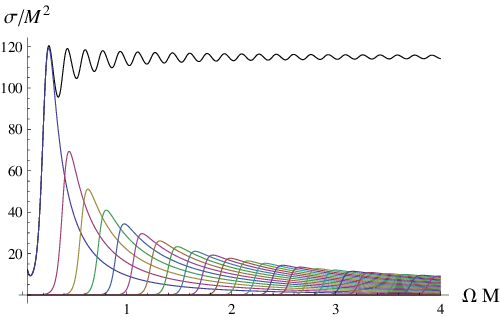}\hfill
\includegraphics[width=0.415\textwidth]{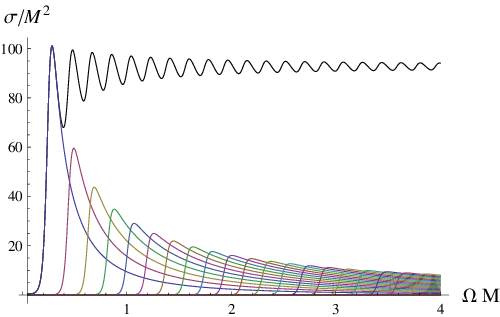}
\end{center}
\noindent\textbf{FIG.~\thefigure.} Effective absorption cross-sections for the electromagnetic sector in generalized Proca backgrounds with $(M,\alpha,\beta,\lambda,c_1)=(1,1,1,0.2,2)$. Left: neutral reference case $Q=0$. Right: near-extreme charged case $Q=1.02$. The colored curves show representative partial absorption cross-sections $\sigma^{(1)}_\ell/M^2$, while the black curves show the total effective absorption cross-sections $\sigma^{(1)}_{\rm abs}/M^2$. Higher multipoles contribute at progressively larger values of $\Omega M$ and produce the oscillatory approach of the total cross-section to its high-frequency regime; increasing $Q$ lowers the overall absorption level while preserving the same qualitative pattern.
\end{widetext}

\section{Thermodynamics}

The same static-patch structure that enters the scattering problem also determines the thermodynamic interpretation of the generalized Proca black holes. In an asymptotically de Sitter spacetime, there is no asymptotically flat region in which the timelike Killing vector can be normalized in the usual way, and the cosmological horizon carries its own temperature. Following the discussion of Schwarzschild--de Sitter radiation in Refs.~\cite{Bousso:1996au,Kanti:2014pha,Pappas:2016ovo,Pappas:2017kam}, we therefore distinguish the bare black-hole temperature, the Bousso--Hawking normalized temperature, and the cosmological-horizon temperature.

For the metric in Eq.~\eqref{eq:metric_function_gbf}, the event and cosmological horizons are the adjacent positive roots $r_h$ and $r_c$ of $f(r)=0$ that bound the static patch. The corresponding bare temperatures are
\begin{equation}
T_0=\frac{|f'(r_h)|}{4\pi},
\qquad
T_c=\frac{|f'(r_c)|}{4\pi}.
\label{eq:bare_temperatures_gbf}
\end{equation}
The normalized black-hole temperature is obtained by rescaling the Killing vector at the geodesic static observer $r_0$, where the attraction of the black-hole horizon and the repulsion of the cosmological horizon balance. In the present coordinates, this point is determined by
\begin{equation}
f'(r_0)=0,
\qquad r_h<r_0<r_c,
\end{equation}
and the Bousso--Hawking temperature is
\begin{equation}
T_{\rm BH}=\frac{T_0}{\sqrt{f(r_0)}}.
\label{eq:bousso_hawking_temperature_gbf}
\end{equation}
As in the Schwarzschild--de Sitter case, one may also define effective temperatures that combine the black-hole and cosmological horizons. For orientation, we quote the positive harmonic-type combination
\begin{equation}
T_{\rm eff}^{(+)}=\frac{1}{T_0^{-1}+T_c^{-1}}
=\frac{T_0T_c}{T_0+T_c},
\end{equation}
however, in the numerical illustrations below, we focus on $T_0$, $T_c$, and $T_{\rm BH}$ because they are the most direct inputs to the Hawking spectra.

\begin{figure*}[t]
\centering
\includegraphics[width=0.98\textwidth]{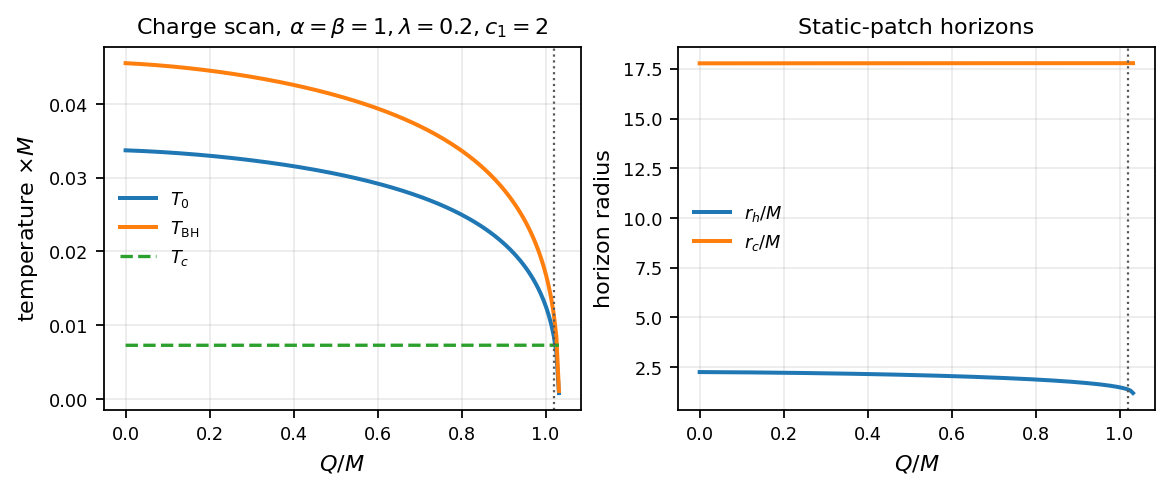}
\caption{
Thermodynamic quantities along the charged generalized Proca family with 
$(M,\alpha,\beta,\lambda,c_1)=(1,1,1,0.2,2)$. 
Left: bare black-hole temperature $T_0$, Bousso--Hawking normalized temperature $T_{\rm BH}$, and cosmological-horizon temperature $T_c$. 
Right: event-horizon and cosmological-horizon radii. 
The vertical dotted line marks the near-extreme case $Q=1.02$ used in Fig.~\ref{fig:acs-em-charge-comparison}.
}
\label{fig:proca-temperature-charge}
\end{figure*}

Figure~\ref{fig:proca-temperature-charge} shows the temperature profile along the charged family with $(M,\alpha,\beta,\lambda,c_1)=(1,1,1,0.2,2)$. The neutral point has $(r_h,r_c)\simeq(2.248,17.776)$ and $(T_0,T_{\rm BH},T_c)\simeq(0.0337,0.0455,0.00728)$. Increasing $Q$ drives the event horizon inward and lowers both black-hole temperatures, while the cosmological horizon and $T_c$ remain almost unchanged. Near the high-charge edge represented by $Q=1.02$, the black-hole temperatures drop to $(T_0,T_{\rm BH})\simeq(0.00845,0.0114)$, explaining why the corresponding absorption and emission channels are strongly suppressed.

The dependence on the couplings is summarized in Fig.~\ref{fig:proca-temperature-couplings} for the neutral family. Varying $\alpha$ mostly changes the size of the static patch: once the very compact end of the scan is avoided, increasing $\alpha$ mildly raises $T_0$ but lowers $T_{\rm BH}$ and $T_c$. The parameter $\beta$ has a stronger and nonmonotonic influence on the normalized temperature because it changes both the near-horizon slope and the normalization point. Increasing $\lambda$ or $c_1$ enlarges the effective de Sitter contribution and raises the cosmological-horizon temperature, while the bare event-horizon temperature decreases. The normalized temperature remains larger than $T_0$ throughout the displayed range, as expected from the factor $1/\sqrt{f(r_0)}$.

\begin{widetext}
\refstepcounter{figure}\label{fig:proca-temperature-couplings}
\begin{center}
\includegraphics[width=0.98\textwidth]{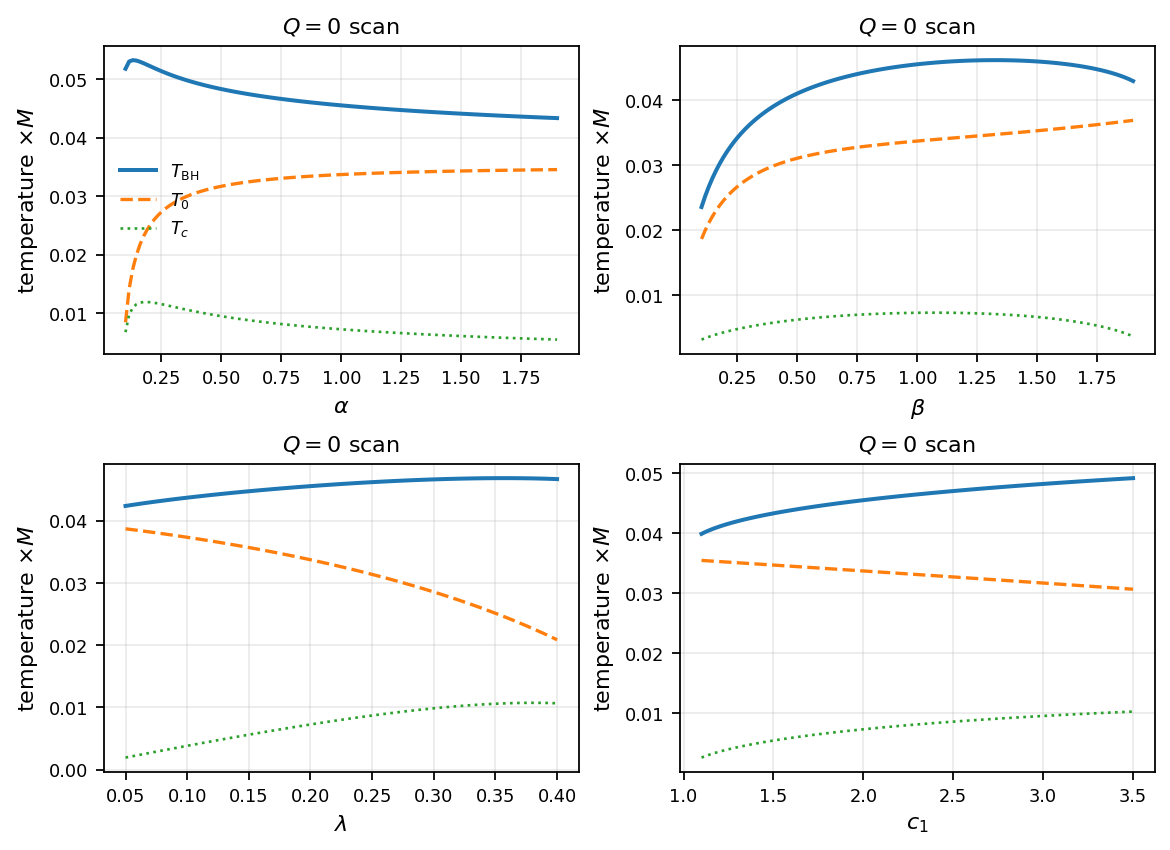}
\end{center}
\noindent\textbf{FIG.~\thefigure.} Temperature scans for the neutral generalized Proca background. In each panel one coupling is varied while the remaining parameters are fixed at $(M,Q,\alpha,\beta,\lambda,c_1)=(1,0,1,1,0.2,2)$. The curves show the normalized black-hole temperature $T_{\rm BH}$, the bare black-hole temperature $T_0$, and the cosmological-horizon temperature $T_c$.
\end{widetext}

\section{Conclusions}

We have analyzed grey-body factors and effective absorption cross-sections for massive scalar, electromagnetic, and massless Dirac fields in the asymptotically de Sitter black holes in generalized Proca theory. The de Sitter behavior considered here is not introduced by adding a bare cosmological constant; it is an effective asymptotic scale generated by the vector sector that also supports the primary hair. This makes the event and cosmological horizons the natural boundaries of the scattering problem and ties the transmission properties directly to the Proca couplings.

For the three test sectors, the radial dynamics reduces to Schr\"odinger-type barrier problems whose potentials vanish at both horizons. The mass of the scalar field is the most sensitive to field parameters, because the scalar mass raises and broadens the barrier in addition to the curvature contribution. As a result, increasing $\mu$ suppresses transmission at fixed frequency and shifts the onset of efficient propagation to larger $\omega$. The electromagnetic and Dirac sectors provide cleaner probes of the background itself: changing the Proca parameters changes the height and width of the barrier, and the grey-body factor follows the expected barrier-penetration pattern.

The sixth-order WKB calculation and the quasinormal-mode reconstruction give consistent transmission curves over most of the plotted frequency range. Their differences are concentrated near the transition region where the grey-body factor rises from strong reflection to strong transmission, precisely where the local barrier-top approximation is most delicate. In the electromagnetic scans, increasing $\alpha$ or $\beta$ hardens the barrier and delays transmission, whereas increasing $c_1$ softens the barrier and moves the transmission window to lower frequencies. In the Dirac sector, increasing the primary-hair parameter $Q$ suppresses the grey-body factor up to higher frequencies.

The effective absorption cross-sections inherit the same physics because they are built directly from the partial transmission probabilities. In the representative electromagnetic comparison, the near-extreme-charged background with $Q=1.02$ has fewer dominant partial-wave peaks and a lower total absorption level than the neutral case. Thus, the primary hair does not merely shift the positions of the horizons; it also reduces the effective capture efficiency of radiation entering from the cosmological-horizon side of the static patch.

These results show that grey-body factors provide a useful complement to quasinormal spectra in effective de Sitter generalized Proca black holes. The next step could be a full direct-scattering computation of $\Gamma_n^{(s)}(\omega)$ and intensities of Hawking radiation for all three sectors across the parameter space, followed by complete Bose--Einstein and Fermi--Dirac emission sums. 

\section*{Declaration of Competing Interest}
The authors declare that they have no known competing financial interests or personal relationships that could have appeared to influence the work reported in this paper.

\section*{Data Availability}
No data was used for the research described in the article.

\begin{acknowledgments}
B. C. L. is grateful to the Excellence project FoS UHK 2205/2025-2026 for the financial support.
\end{acknowledgments}

\bibliographystyle{apsrev4-1}
\bibliography{ProcaGreyBodyFactors}

\end{document}